\documentclass[floatfix,aps,amsmath,nofootinbib,onecolumn,10pt]{revtex4}

\usepackage{graphicx}
\usepackage{bm}
\usepackage{rotating}
\usepackage{array}

\def\({\left(}
\def\){\right)}
\def\[{\left[}
\def\]{\right]}

\def\e{\begin{equation}}
\def\q{\end{equation}}
\def\m{\begin{eqnarray}}
\def\n{\end{eqnarray}}

\begin{document}

\title{Does the Reionization Model Influence the Constraints on Dark Matter Decay or Annihilation?}

\author{Lu Chen$^{1}$ \footnote{chenlu@sdnu.edu.cn} and Ke Wang$^{2}$ \footnote{wangkey@lzu.edu.cn}$^,$ \footnote{Corresponding author}}
\affiliation{$^1$ School of Physics and Electronics, \\Shandong Normal University, Jinan 250014, China\\
$^2$ Institute of Theoretical Physics \& Research Center of Gravitation,\\ Lanzhou University, Lanzhou 730000, China\\} 
\date{\today}

\begin{abstract}

If dark matter decay or annihilate, a large amount of energy and particles would be released into the cosmic plasma. Therefore, they could modify the thermal and ionization history of our universe, then leave footprints on the cosmic microwave background power spectra.
In this paper, we take dark matter annihilation as an example and investigate whether different reionization models influence the constraints on dark matter  annihilation.
We consider the ionization history including both dark matter annihilation and star formation, then put constraints on DM annihilation.
Combining the latest Planck data, BAO data, SNIa measurement, $Q_\text{HII}$ constraints from observations of quasars, as well as the star formation rate density from UV and IR data, the optical depth is $\tau=0.0571^{+0.0005}_{-0.0006}$ at 68$\%$C.L. and the upper limit of $\epsilon_0 f_d$ reads $2.7765\times 10^{-24}$ at 95$\%$C.L..
By comparison, we also constrain dark matter annihilation in the instantaneous reionization model from the same data combination except the $Q_\text{HII}$ constraints and star formation rate density. We get $\tau=0.0559^{+0.0069}_{-0.0076}$ at 68$\%$C.L. and the upper limit of $\epsilon_0 f_d$ is $2.8468\times 10^{-24}$ at 95$\%$C.L..
This indicates various reionization models have little influence ($\lesssim 2.5\%$) on constraining parameters of dark matter decay or annihilation.

\end{abstract}

\pacs{???}

\maketitle


\section{Introduction} 
\label{sec:int}

The existence of dark matter (DM) has been confirmed by cosmological observations, for example, gravitational lensing of clusters and galaxies~\cite{Hoekstra:2002nf,Koopmans:2002qh,Metcalf:2003sz,Moustakas:2002iz}, the large-scale structure formation~\cite{Davis:1985rj,Navarro:1995iw,Navarro:2003ew,Davis:1992ui,Blumenthal:1984bp}, rotation curves of galaxies~\cite{vandenBosch:2000rza,Begeman:1991iy,deBlok:2001hbg}, the cosmic microwave background (CMB)~\cite{Ade:2015xua,Aghanim:2018eyx} and so on~\cite{Jungman:1995df,Bertone:2004pz,Clowe:2006eq,Peter:2012rz}. 
However, its nature is still a puzzle after decades of researches and observations~\cite{Lewin:1995rx,Gaskins:2016cha,Kahlhoefer:2017dnp,Freese:2012xd,Cirelli:2012tf}.
Many theories predict that DM particles can decay or annihilate into standard model particles (e.g. photons, protons, electrons and so on) which are well known in particle physics~\cite{Geringer-Sameth:2014yza,Steigman:2012nb,Gondolo:1999ef,Galli:2009zc,Galli:2011rz}.
It is clear that the thermal and ionization history of our universe would be influenced because annihilating or decay DM release extra particles and energy into the cosmic plasma~\cite{Zhang:2006fr,Dvorkin:2013cga,Weniger:2013hja,Chluba:2009uv}.
If it occurs during the epoch of recombination, DM decay or annihilation would lead to extra ionizations and excitations, as well as the increase of plasma temperature. As a result, recombination would be modified.
Moreover, DM decay or annihilation occurred during the reionization epoch could 
be an extra ionized source besides star formation.
Eventually, the whole process would leave footprints on the CMB power spectra and provide a method to constrain DM using CMB observations.
Actually, the recombination redshift $z_* \sim 1089$ is constrained well from observational data in various recombination models~\cite{Ade:2015xua,Aghanim:2018eyx}.
On the other hand, observations of the Gunn-Peterson effect in the spectra of high-redshift quasars reveal that the reionization epoch of our universe ended before redshift $z\sim 6 $~\cite{Robertson:2015uda,Gruppioni:2011vi,Chornock:2013una,Madau:2014bja}. But the history during $z\sim 6-1089$ is still under debate, especially the onset and process of reionization. 

Although previous works indicate that DM decay or annihilation contributed much less ionized photons than stars and galaxies during the epoch of reionization~\cite{Lopez-Honorez:2013cua,Carucci:2018tzf,Liu:2016cnk,Mapelli:2006ej,Hazra:2018eib,Paoletti:2020ndu}, that occurred during recombination has the ability to influence the reionization history by providing the initial conditions. 
The most popular reionization model is the instantaneous reionization model with a simple parameterization of hyperbolic tangent function $\tanh$ (named the TANH model hereafter) $\footnote{\ \ \ In the TANH model, the ionized fraction $x_e=N_e/N_\text{H} \simeq x_{\text{p}}+x_{\text{HeII}}$ is assumed to be~\cite{Lewis:2008wr} \\ \m\nonumber x_e=\dfrac{1+N_\text{He}/N_\text{H}}{2}\[1+\tanh\(\dfrac{y(z_\text{re})-y(z)}{\Delta y}\)\] ,\n \\where $y(z)=(1+z)^{3/2}$, $\Delta y = \dfrac{3}{2}(1+z_{\text{re}})^{1/2}\Delta z$, $\Delta z =0.5$. And the optical depth is \\ \m\nonumber \tau(z)=\int_0^z x_e N_{\text{H}} \sigma_{\text{T}} H^{-1}(z')(1+z')^{-1} dz'.\n}$. Planck Collaboration gave the reionization optical depth $\tau = 0.054 \pm 0.007$ at 68$\%$C.L. in 2018 (hereafter Planck 2018)~\cite{Aghanim:2018eyx}.
An extended model named Poly-reion model~\cite{Hazra:2017gtx} has also been well studied. The ionized fraction is patameterized by a polynomial function and higher values of $\tau$ are acquired, which are similar with values obtained using the principle-component analysis (PCA)~\cite{Heinrich:2016ojb}.
However, there is another much more direct approach to study the cosmic reionization history via star formation history (SFH). Since early stars and galaxies are considered as ionized sources of reionization~\cite{Robertson:2015uda,Khusanova:2019cxr,Madau:2014bja,Robertson:2010an,Robertson:2013bq}, we can reconsider the reionization history using star formation rate (SFR) density from UV and IR data~\cite{Madau:2014bja} at lower redshifts directly (named the SFH model hereafter).
In this paper, we take DM annihilation into consideration and put constraints on parameters in the TANH and SFH reionization models from observational data. 

This paper is organized as follows.
In section~\ref{sec:his}, the ionization history of universe is reconsidered including DM annihilation and SFR rate. 
In subsection~\ref{sec:dma}, we review the modified ionized fraction after taking DM annihilation into account.
In subsection~\ref{sec:sfr}, the SFR density and its influence on the ionization of the intergalactic medium (IGM) during the epoch of reionization are presented.
In subsection~\ref{sec:tau}, we give the required equations for calculating the optical depth.
Our constraints on DM annihilation are shown in section~\ref{sec:res}.
In the SFH model, we make use of the data combination including CMB measurement from Planck 2018~\cite{Aghanim:2018eyx}, SFR density from UV data and IR data~\cite{Madau:2014bja}, $Q_\text{HII}$ constraints from observations of quasars~\cite{Bouwens:2015vha,Fan:2005es,Schenker:2014tda}, baryon acoustic oscillation (BAO) data~\cite{Beutler:2011hx,Ross:2014qpa,Alam:2016hwk}, as well as the measurement of Type Ia supernova~\cite{Scolnic:2017caz}. 
In the TANH model, the same data combination excluding SFR density is used.
Finally, a brief summary is included in section~\ref{sec:sum}.

\section{Ionization History}
\label{sec:his}


\subsection{Effects of Annihilating Dark Matter on Recombination and Reionization}
\label{sec:dma}

In this subsection, we give a brief summary of the ionized fraction related to DM annihilation~\cite{Chluba:2009uv}.
Considering self-annihilating DM particles $\chi$ and their anti-particles $\bar{\chi}$, the total energy rate released via annihilation per unit volume usually depends on the involved annihilation channels, mass of DM particles $M_{\chi}=M_{\bar{\chi}}$ and their averaged cross-section $<\sigma v>$. But only a fraction of them can be deposited into the cosmic plasma. The energy deposition efficiency $f_\text{d}$ is related to redshift $z$ and cosmological models. In brief, the total rate of energy deposited to the cosmic plasma per unit volume is given by
\m
\dfrac{dE_\text{d}}{dt}\Big{|}_{\chi \bar{\chi}}=\epsilon_0 f_\text{d} N_\text{H}(z)(1+z)^3 \text{eV} \cdot \text{s}^{-1},
\n 
where $N_\text{H}(z)$ is the comoving number density of hydrogen nuclei in our universe and $\epsilon_0$ is a dimensionless parameter given by
\m
\epsilon_0 = 1.5 \times 10^{-24} \Big{[} \dfrac{M_{\chi}c^2}{100 \text{GeV}} \Big{]}^{-1} \Big{[} \dfrac{\Omega_{\chi}h^2}{0.13} \Big{]}^{2} \Big{[} \dfrac{<\sigma v>}{3\times 10^{-26}\text{cm}^3  \cdot  \text{s}^{-1}} \Big{]},
\n
where $c$ is the speed of light, $\Omega_{\chi} h^2$ is the physical density of DM today. In this paper, $\epsilon_0 f_\text{d}$ is regarded as a constant roughly.

The energy absorbed by the cosmic plasma will heat, ionize and excite atoms.
To measure the effects of annihilating DM on the ionization history, the evolution equations for the net ionization rate from ground states of neutral hydrogen and helium results from DM annihilation, as well as the plasma temperature evolution equation, are written as~\cite{Chluba:2009uv,Lopez-Honorez:2013cua}
\begin{eqnarray}
\label{eq:x}
     \dfrac{d x_\text{p}}{dt}\Big{|}_{\chi \bar{\chi}} &=&- \dfrac{1}{N_{\text{H}}[1+f_{\text{He}}]} \dfrac{g_\text{p}(z)}{E^{\text{HI}}_{\text{ion}}}  \dfrac{dE_d}{dt}\Big{|}_{\chi \bar{\chi}} \,,\nonumber\\
     \dfrac{d x_\text{HeII}}{dt}\Big{|}_{\chi \bar{\chi}} &=&- \dfrac{f_{\text{He}}}{N_{\text{H}}[1+f_{\text{He}}]} \dfrac{g_{\text{He}}(z)}{E^{\text{HeI}}_{\text{ion}}} \dfrac{dE_d}{dt}\Big{|}_{\chi \bar{\chi}} \,,\nonumber\\
     \dfrac{d T_\text{m}}{dt}\Big{|}_{\chi \bar{\chi}} &=&  \dfrac{1}{N_{\text{H}}[1+f_{\text{He}}+x_\text{e}(z)]} \dfrac{2g_\text{h}(z)}{3k_{\text{B}}}  \dfrac{dE_d}{dt}\Big{|}_{\chi \bar{\chi}} \,,\nonumber\\
 \end{eqnarray}  
where $ x_\text{p}\equiv N_\text{p}/N_{\text{H}}$, $ x_\text{HeII}\equiv N_\text{HeII}/N_{\text{H}}$ are the ionized fraction from the ground states of hydrogen and helium related to DM annihilation respectively, $f_{\text{He}}\equiv N_{\text{He}}/N_{\text{H}}\simeq 0.08$ is the fraction of helium, $E^{\text{HI}}_{\text{ion}}=13.6\ \text{eV}$ and $E^{\text{HeI}}_{\text{ion}}=24.6 \ \text{eV}$ are the ionization potentials of hydrogen and helium respectively, $k_{\text{B}}$ is the Boltzmann constant. $g_{\text{h}}$, $g_\text{p}$ and $g_{\text{He}}$ represent the partial contributions of the total deposited energy going into heating the plasma, ionizations of hydrogen and helium. Here we use the following approximations,
\begin{eqnarray}
\label{eq:g}
     g_{\text{p}} &\approx & \dfrac{1-x_{\text{p}}(z)}{3} \,,\nonumber\\
     g_{\text{He}} &\approx & \dfrac{1-Z                                                             _{\text{HeII}}(z)}{3} \,,\nonumber\\
     g_{\text{h}} &\approx & \dfrac{1}{3} \( 1+\dfrac{2x_{\text{e}}(z)}{1+f_{\text{He}}} \) \,,\nonumber\\
 \end{eqnarray}  
where $Z_{\text{HeII}}=N_\text{HeII}/N_{\text{He}}$ is the fraction of singly ionized helium atoms.
For simplification, $T_\text{m}$ is assumed to be a constant during the epoch of reionization, which is different from the redshift-dependent temperature $T_\text{m}$ during recombination . 
This is reasonable because the duration of recombination is much longer than reionization and the change of $T_\text{m}$ during reionization epoch is small enough to be ignored.
Notice that we have ignored the contribution of excitations because electrons are tend to stay at lower levels and it has a weak influence relative to ionizations~\cite{Lopez-Honorez:2013cua,Galli:2011rz,Hutsi:2011vx}.


\subsection{Effects of Star Formation on Reionization}
\label{sec:sfr}

On the other hand, the mainstream considers early star formation of galaxies most likely supplies the ionizing photons for the reionization of our universe~\cite{Madau:2014bja,Robertson:2015uda}. Naturally, SFR involves significant information on the onset and process of reionization.
We use the following parameterization of cosmic SFR density $\rho_\text{SFR}$ (in units of $\text{M}_{\odot} \cdot \text{yr}^{-1} \cdot \text{Mpc}^{-3}$)~\cite{Madau:2014bja,Bowman:2018yin},
\m
\rho_{\text{SFR}} = \dfrac{a_{\text{p}} (1+z)^{b_{\text{p}}}}{1+[(1+z)/c_{\text{p}}]^{d_{\text{p}}}} - \dfrac{a_{\text{p}} (1+22)^{b_{\text{p}}}}{1+[(1+22)/c_{\text{p}}]^{d_{\text{p}}}},
\n
with $a_\text{p}$, $b_\text{p}$, $c_\text{p}$ and $d_\text{p}$ are the four coefficients which ought to be determined by observational data.
The second term results from our assumption that there's no star formation before $z \sim 22$.
We choose the value of 22 because there is almost no 21 cm absorption signal before $z \sim 22$ as the figure of best-fit 21 cm absorption profiles shows in \cite{Bowman:2018yin}.
This parameterization not only can provide the bump at about $z\sim 2$, but also behaves as $\log \rho_\text{SFR} \rightarrow -\infty $ near the starting redshift of reionization.
Therefore, $\log \rho_\text{SFR}(z\gtrsim 10)$ performs a rapid downtrend  which is consistent with Fig.~8 of \cite{Ishigaki_2018}.
If Lyman continuum photons released from star forming galaxies play a dominant role in the reionization, their influence on the ionized rate depends on the photon production rate $\xi_\text{ion}$ and the escaping rate $ f_\text{esc}$ of galaxies besides $\rho_\text{SFR}$, which can be written as the growth rate of number density for ionized photons,
\m
\label{eq:sfr}
\dot{N}_\text{ion} = f_\text{esc} \xi_\text{ion} \rho_\text{SFR},
\n
with the overdot indicates the differentiation of cosmic time $t$. Here, we adopt the fiducial values $f_\text{esc}=0.2$ and $\log_{10} \xi_\text{ion} = 53.14\  [\text{Lyc} \cdot \text{photons} \cdot  \text{s}^{-1}\cdot  \text{M}_{\odot}^{-1}\cdot  \text{yr}]$.


\subsection{Thomson Optical Depth}
\label{sec:tau}
The Tomson optical depth is given by
\m
\tau(z) = c  \sigma_{\text{T}} \int_0^z (1+f_{\text{He}}) Q_{\text{HII}}(z')N_{\text{H}}(z') (1+z')^{-1} H^{-1}(z')dz',
\n
where $c$ is the speed of light, $\sigma_\text{T}$ is the Thomson scattering cross section, $H(z)$ is the redshift-dependent Hubble parameter. And $Q_{\text{HII}}$ indicates the ionized fraction (equivalent to $x_\text{p}$ physically) which can be calculated by evolving the differential equation
\m
\label{eq:q}
\dot{Q}_{\text{HII}} =  -\dfrac{Q_{\text{HII}}}{t_{\text{rec}}} + \dfrac{d x_\text{p}}{dt}\Big{|}_{\chi \bar{\chi}} + \dfrac{\dot{N}_{\text{ion}}}{N_{\text{H}}} .
\n
Here $t_{\text{rec}}$ is the recombination time as follows~\cite{Bouwens:2015vha}
\m
t_{\text{rec}} = 0.88\  \text{Gyr} \(\dfrac{1+z}{7}\)^{-3} \(\dfrac{T_0}{2\times 10^4 \text{K}}\)^{-0.7} \( \dfrac{C_{\text{HII}}}{3}\)^{-1},
\n
where $T_0 \sim 2\times 10^4$ K is the temperature of the ionizing hydrogen gas and $C_{\text{HII}} \sim 3$ is the clumping factor of ionized hydrogen.
The second term of Eq.~\ref{eq:q} represents the effect of DM annihilation and the last term represents the effect of SFR.
Note we have ignored the contribution of He ionization in the second equation of Eq.\ref{eq:x}.
The initial value of $Q_{\text{HII}}$ at the start of reionization depends on DM annihilation occurred during the recombination epoch and that is why we can constrain $\epsilon_0 f_\text{d}$.


\section{Results}
\label{sec:res} 
 
In this section, we use the observational data to reconstruct the ionization history of our universe and put constraints on DM annihilation.
CAMB~\cite{Lewis:1999bs} and CosmoRec~\cite{Shaw:2011ez} are used to calculate the optical depth and CMB power spectra. Then the Markov Chain Monte Carlo sampler---CosmoMC~\cite{Lewis:2002ah,Lewis:2013hha} is applied to constrain the free parameters using the maximum likelihood determination.
In the SFH model, we refer to the data combination of Planck 2018 TT,TE,EE$+$lowE$+$lensing~\cite{Aghanim:2018eyx}, baryon acoustic oscillation (BAO) data at redshifts $z = 0.106,0.15,0.32,0.57,1.52$ (named 6dFGS~\cite{Beutler:2011hx}, MGS~\cite{Ross:2014qpa}, DR12~\cite{Alam:2016hwk} respectively), the latest SNIa measurement (named Pantheon)~\cite{Scolnic:2017caz}, $Q_\text{HII}$ constraints between $5.0\leq z \leq 8.0$ from observations of Gunn-Peterson optical depth and prevalence of Ly-$\alpha$ emission in galaxies~\cite{Bouwens:2015vha,Fan:2005es,Schenker:2014tda},as well as the SFR density from UV and IR data~\cite{Madau:2014bja}.
Notice that the SFR data used here extend to $z\sim 8$. Actually, \cite{Bouwens:2015vha,Ishigaki_2018} provide UV data extending to $z\sim 10$, but the uncertainties of the data are large between $8<z<10$. So our choice would not impact the results significantly. 
There are ten free parameters in this model, $\{\Omega_{\text{b}}h^2, \Omega_{\text{c}}h^2, 100\theta_{\text{MC}}, \text{ln} \(10^{10} A_\text{s}\), n_{\text{s}}, 10^{-23}\epsilon_0f_\text{d},a_\text{p},b_\text{p},c_\text{p},d_\text{p} \}$. 
Here five of them are parameters from the $\Lambda$CDM model: $\Omega_\text{b}h^2$ and $\Omega_\text{c}h^2$ are the density of baryons and cold dark matter today respectively, $100\theta_\text{MC}$ is 100 times the ratio of angular diameter distance to the large scale structure sound horizon, $A_\text{s}$ is the amplitude of the power spectrum of primordial curvature perturbations, and $n_\text{s}$ is the scalar spectrum index. Besides, $\epsilon_0 f_\text{d}$ and $a_\text{p},b_\text{p},c_\text{p},d_\text{p}$ are parameters of DM annihilation and SFR density as described in section~\ref{sec:his}.
For comparison, the TANH model with previous data combination except the SFR density data has been run, too. In this model, we have seven free parameters $\{\Omega_{\text{b}}h^2, \Omega_{\text{c}}h^2, 100\theta_{\text{MC}}, \text{ln} \(10^{10} A_\text{s}\), n_{\text{s}}, \tau, 10^{-23}\epsilon_0f_\text{d} \}$.
Moreover, we have to mention that the optical depth $\tau$ is a free parameter in the TANH model, but it's a derived parameter in the SFH model because SFR density data are used to put constraints on the reionization epoch directly, replacing the parameterization in the TANH model.

\begin{figure}[]
\begin{center}
\includegraphics[scale=0.3]{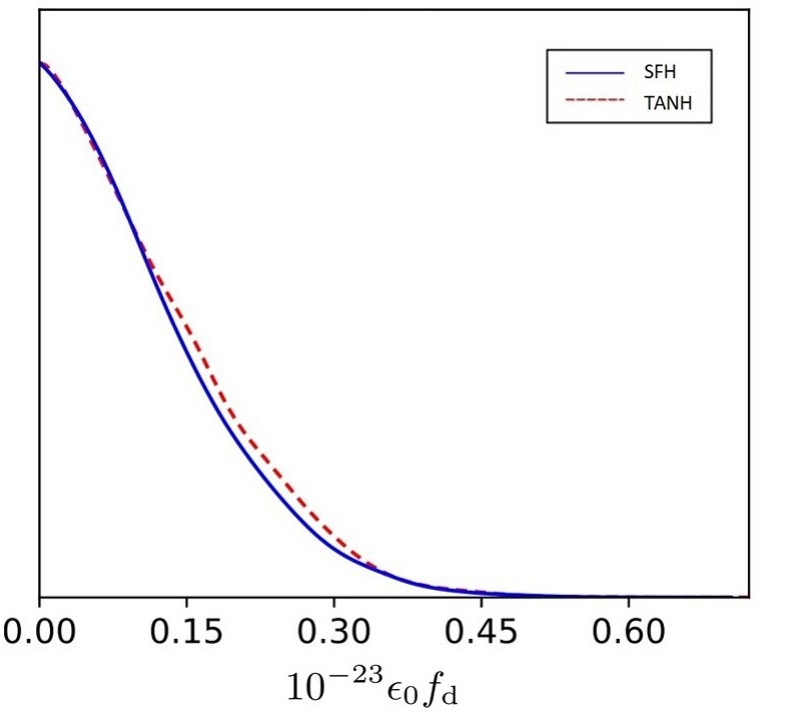}
\end{center}
\caption{The likelihood of $10^{-23}\epsilon_0 f_\text{d}$ in the SFH and TANH model. The blue solid curve indicates the SFH model and the red dashed curve indicates the TANH model.}
\label{fig:1d}
\end{figure}

\begin{figure}[]
\begin{center}
\includegraphics[scale=0.3]{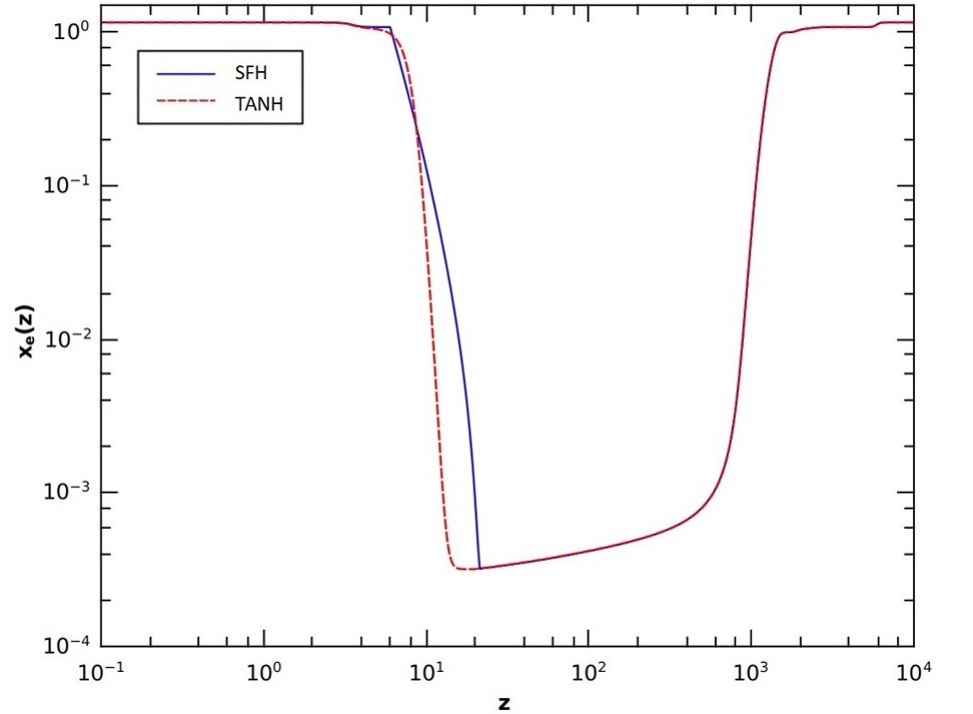}
\end{center}
\caption{The evolution of ionized fraction $x_e(z)$ in the SFH and TANH model. The blue solid curve illustrate the SFH model and the red dashed curve corresponds to the TANH model. Note that the values of $\epsilon_0 f_\text{d}$ are set to the upper limits at 95$\%$C.L. and we set the other necessary parameters to their mean values in Tab.~I.}
\label{fig:xe}
\end{figure}

Our results are summarized in Tab.~I. We have listed the 68$\%$ limits of the free parameters and necessary derived parameters in these two models.
The upper limit of $\epsilon_0 f_\text{d}$ is $1.3309\times 10^{-24}$ at 68$\%$ C.L. and $ 2.7765\times 10^{-24}$ at 95$\%$ C.L. in the SFH model, while it reads $ 1.4253\times 10^{-24}$ at 68$\%$ C.L. and $ 2.8468\times 10^{-24}$ at 95$\%$ C.L. in the TANH model. 
The constraint on $\epsilon_0 f_\text{d}$ improves 6.6$\%$ at 68$\%$ C.L. and 2.5$\%$ at 95$\%$ C.L..
The likelihood is shown vividly in Fig.~\ref{fig:1d}.
Our results reveal modifying reionization epoch has little influence on constraining DM annihilation.
In the SFH model, the optical depth $\tau$ is $0.0571^{+0.0005}_{-0.0006}$ at 68$\%$C.L.. It is in consistent with the value of $0.0559_{-0.0076}^{+0.0069}$ in the TANH model, as well as $0.054 \pm 0.007$ in the $\Lambda$CDM model released by Planck Collaboration in 2018~\cite{Aghanim:2018eyx}. 
The mean values of $\tau$ do not vary almostly, leading to similar constraints on $\epsilon_0 f_\text{d}$. This is because they are related directly as shown in Sec.~\ref{sec:his}. 
However, the error bar of $\tau$ is reduced by around an order of magnitude in the SFH model.
Notice previous reionization analysis including both CMB and SFR data gave uncertainties of $\tau$ about $\sigma_{\tau} \sim 0.001-0.002$~\cite{Gorce:2017glg,Krishak:2021fxp}.
Our error bars are about half smaller on account of the $Q_\text{HII}$ constraints of Gunn-Peterson optical depth between $5.03<z<5.85$, which provide strict restrictions of $Q_\text{HII}$ with uncertainties of order $\sigma \sim 0.00001-0.00005$~\cite{Bouwens:2015vha,Fan:2005es}.
Besides, the ionized fraction $x_e(z)$ in these two models are illustrated in Fig.~\ref{fig:xe}. It shows that the reionization started at about $z \sim 20$ in the SFH model, which is higher than the TANH model.

\begin{table}
\label{tb:result}
\caption{The 68$\%$ limits for the cosmological parameters in the SFH and TANH model. Note we give the upper limits of $10^{-23}\epsilon_0f_\text{d}$ at 95$\%$C.L.. }
\begin{tabular}{p{3.5 cm}<{\centering}|p{4.5cm}<{\centering} p{4.5cm}<{\centering}   }
\hline
                  & SFH                      & TANH \\
\hline
$\Omega_\text{b} h^{2}$& $0.02240\pm 0.00013$        & $0.02245_{-0.00013}^{+0.00014}$ \\
$\Omega_\text{c} h^{2}$& $0.1192\pm 0.0008$          & $0.1193\pm 0.0009$\\
$100\theta_\text{MC}$&$1.04094_{-0.00029}^{+0.00030}$&$1.04095_{-0.00029}^{+0.00030}$ \\
$\ln(10^{10}\text{A}_\text{s})$&$3.0529_{-0.0071}^{+0.0063}$& $3.0509\pm 0.0144$\\
$n_s$             & $0.9675\pm 0.0036$               & $0.9677\pm 0.0037$ \\
\hline
$10^{-23}\epsilon_0f_\text{d} (95\% $ C.L.) &$<0.27765$                   &$<0.28468$\\
$a_{\text{p}}$[$\text{M}_{\odot} \cdot \text{yr}^{-1} \cdot \text{Mpc}^{-3}$]   & $0.01769_{-0.00069}^{+0.00068}$ &-\\
 $b_{\text{p}}$   &   $2.967\pm0.118 $              &- \\
 $c_{\text{p}}$   &   $2.555_{-0.100}^{+0.087}$     &- \\
 $d_{\text{p}}$   & $5.131_{-0.095}^{+0.096}$       &- \\
 \hline
 $\tau$           &   $0.0571_{-0.0006}^{+0.0005} $ &$0.0559_{-0.0076}^{+0.0069}$      \\
 $10^{-4}Q_\text{HII}(z=22)$&$2.666_{-0.511}^{+0.220}$   &- \\
$H_0$[km $\cdot$ s$^{-1}$ $\cdot$ Mpc$^{-1}$]&$67.66_{-0.38}^{+0.39}$&$67.69\pm 0.42$  \\
 \hline
\end{tabular}
\end{table}

\section{Summary and discussion}
\label{sec:sum}

In this paper, we investigate whether the reionization model influences constraining parameters of DM annihilation or decay. Take DM annihilation as an example.
Energy released by annihilating DM would be absorbed by the cosmic plasma, resulting extra heating, ionization and excitation of gas. Therefore, both the recombination and reionization history of our universe are influenced, leaving footprints on the CMB power spectra. 
On the other hand, early star forming galaxies are thought to be sources of reionization epoch. This provides a way to reconstruct the ionization history with the SFR density directly.
Combining the measurement of Planck 2018, BAO data, PANTHEON samples, $Q_\text{HII}$ constraints, as well as the SFR density from UV and IR data, we put constraints on the ionization history with DM annihilation and star formation by performing a maximum likelihood determination with CAMB$+$CosmoMC packages.
The upper limits of $\epsilon_0 f_d $ show $1.3309\times 10^{-24}$ at 68$\%$ C.L. and $ 2.7765\times 10^{-24}$ at 95$\%$ C.L. in this model.
By comparison, we also research the instantaneous reionization model including DM annihilation from the same data combination except SFR density data and $Q_\text{HII}$ constraints.
We get similar constraints and $\epsilon_0 f_d $ reads $1.4253\times 10^{-24}$ at 68$\%$ C.L. and $ 2.8468\times 10^{-24}$ at 95$\%$ C.L..
So the reionization models have little influence ($\lesssim 2.5\%$) on constraining DM annihilation.
Similarly, DM decay has analogous influence on ionization history of our universe and the decay rate plays a similar role as $\epsilon_0 f_\text{d}$.
Therefore, we arrive at a conclusion that various reionization models have little effects on DM decay or annihilation.
Besides, the optical depth is $\tau=0.0571^{+0.0005}_{-0.0006}$ at 68$\%$ in the SFH model, improving significantly than $\tau=0.0559^{+0.0069}_{-0.0076}$ in the TANH model.

In our models, we have made some simplifications.
We use the tipical values of parameters, such as the clumping factor of ionized hydrogen $C_\text{HII} \sim 3$, the escaping rate $f_\text{esc}=0.2$ and so on. Actually, other comparable values or redshift-varying functions of them are also acceptable~\cite{Ouchi:2009ys,Carucci:2018tzf,Robertson:2010an}. However, we do not think these comparable values would change our conclusion significantly.
More discussion about complicated models can be found in \cite{Carucci:2018tzf,Robertson:2013bq,Robertson:2010an,Gorce:2017glg}.
We also assume $\epsilon_0 f_d$ to be a constant though it varies with redshift, cosmological models, as well as DM annihilation channels.
The plasma temperature evolution and excitation during the epoch of reionization are also small enough to be ignored.
Moreover, DM annihilation in halos may also influence the reionization history of our universe~\cite{Lopez-Honorez:2013cua,Liu:2016cnk}. 
But we have ignored the enhancement of halos in this paper.

\vspace{5mm}
\noindent {\bf Acknowledgments}

We acknowledge the use of HPC Cluster of Tianhe II in National Supercomputing Center in Guangzhou. Ke Wang is supported by grants from NSFC (grant No. 12005084) and grants from the China Manned Space Project with NO. CMS-CSST-2021-B01. Lu Chen is supported by grants from NSFC (grant No. 12105164). This work has also received funding from project ZR2021QA021 supported by Shandong Provincial Natural Science Foundation.



\end{document}